# Post-hatching parental care behaviour and hormonal status in a precocial bird.

Boos, Mathieu[1,2], ZIMMER Cédric[1], Carriere Aurélie[1], Robin Jean-Patrice[1], and Odile Petit[1]

[1] Institut Pluridisciplinaire Hubert Curien, Département Ecologie, Physiologie et Ethologie, UMR 7178 CNRS-ULP, 23 rue Becquerel, 67087 Strasbourg Cedex 2, France. Phone: +33 3 88 10 69 24; Fax: +33 3 88 10 69 06

[2] Naturaconst@ 14 rue principale F-67 270 Wilshausen.
Phone / Fax: +33 3 88 02 26 76.

Correspondence: Odile PETIT, odile.petit@c-strasbourg.fr

## Abstract

In birds, the link between parental care behaviour and prolactin release during incubation persists after hatching in altricial birds, but has never been precisely studied during the whole rearing period in precocial species, such as ducks. The present study aims to understand how changes in parental care after hatching are related to circulating prolactin levels in mallard hens rearing ducklings. Blood was sampled in hens over at least 13 post-hatching weeks and the behaviour of the hens and the ducklings was recorded daily until fledging. Contacts between hens and the ducklings, leadership of the ducklings and gathering of them steadily decreased over post-hatching time. Conversely, resting, preening and agonistic behaviour of hens towards ducklings increased. Plasma prolactin concentrations remained at high levels after hatching and then fell after week 6 when body mass and structural size of the young were close to those of the hen. Parental care behaviour declined linearly with brood age, showed a disruption of the hen-brood bond at week 6 post-hatching and was related to prolactin concentration according to a sigmoid function. Our results suggest that a definite threshold in circulating prolactin is necessary to promote and/or to maintain post-hatching parental care in ducks.

# INTRODUCTION

In birds, parental care behaviour is a key concept of parental investment. According to life-history traits, it is selected for differently among bird species and relies on the trade-off between increasing survival of the offspring while impairing that of the parents for future reproduction (Kear, 1970; Afton and Paulus, 1992). Trivers (1974) has suggested that the parent-offspring conflict is expected to increase during the period of parental care, and offspring are expected to compete with their parents. During incubation, parental care is linked to prolactin secretion in a large number of species (Schradin and Anzenberger, 1999). However, after hatching, although it has been shown that this hormone is involved in the control of parental behaviour in altricial birds (Silverin and Goldsmith, 1984; Hall *et al.*, 1986; Schoech *et al.*, 1996; Lormée *et al.*, 1999), its role in precocial species is more controversial (Goldsmith, 1991). Indeed, depending on the species, either sharp drops or slow decreases of adult prolactin levels after hatching have been reported (Goldsmith and Williams, 1980; Dittami, 1981; Oring *et al.*, 1986; Hall, 1987; Oring *et al.*, 1988). While data for common eiders, *Somateria mollissima,* suggest that a threshold prolactin level should be reached to promote parental care (Criscuolo *et al.*, 2002), the direct link between the hormonal status and the intensity of this behaviour has in fact never been specifically explored.

To investigate such a possible hormonal-behavioural relationship in precocial birds, the mallard duck *Anas platyrhynchos* is a suitable model. In this species, a drop in prolactin plasma concentration has been reported just after hatching (Goldsmith and Williams, 1980; Hall, 1987), while a strong hen-brood attentiveness has been observed at least during the first 2-3 weeks post-hatching, before a complete disruption of the bond in about 6-7 weeks old ducklings (i.e. 2-3 weeks before fledging, Talent *et al.*, 1983; Afton and Paulus, 1992). From these data it could be concluded that if prolactin is still involved in parental care after hatching in such species, plasma levels of prolactin should be correlated with the intensity of the hen-brood bond and be maintained above a threshold value. To check this hypothesis, we performed here the first detailed study of parental care behaviours in mallard females during the rearing of ducklings, from hatching to fledging, in relation to the hormonal status of the females over the same period. To take into account possible effects of the photoperiod on prolactin secretion and nest abandonment (Hector and Goldsmith, 1985; Bluhm, 1992; Lormée *et al.*, 1999; Sockman *et al.*, 2004) the present study was undertaken over spring and summer. Together with prolactin, plasma levels of corticosterone, known to interact with prolactin release (Criscuolo, 2001; for review see Chastel *et al.* 2005) and testosterone (potentially involved in aggressive behaviour, Dittami, 1981) were also measured.

## MATERIALS AND METHODS

*Animals and experimental condition.*

The present study was conducted in 2003 on 7 adult mallard pairs obtained from the registered breeding field station of the Canarderie de la Ronde (Cère la Ronde, France). About 2 months before the nesting period, the pairs were held in separated outdoor pens (12 m²), each including a nest and a 1m² basin provided with clear running water. Each side of the pen (except the top) was opaque to avoid any visual contact between the pens and to prevent any external stress and especially from the observers during periods of behaviour monitoring. After laying and as soon as the hen started incubation, males were removed from the pen. Ducklings hatched between 15 April and

23 June and were continuously held with the hen until mid September, i.e. well after fledging.

The ducks were subjected throughout the study to ambient temperature and natural photoperiods. A balanced commercial food adapted for adults and one specific for ducklings (Sanders Corporation), along with fresh drinking water, were provided *ad libitum*.

*Blood sampling and weighing*

Blood samples were taken from all hens every 3/4 days during at least 13 weeks from hatching, between 10:00 and 12:00 am. Hens were caught in the pen with a net and transported to a nearby room. Within three minutes after capture, blood was collected into a heparinized tube by puncture of the brachial vein, centrifugated and then plasma was frozen at -20°C until analysis. Before being released back to the pen, the hen was weighed (±1g) and the post-nuptial moulting status of the plumage was determined by visual examination. Three scores were distinguished: no moult, body moult (feathers along side and flanks), moult of wing feathers. Body mass was used as a condition index. No size corrections were applied since structural measurements do not significantly explain body reserves in hen mallards and can be misleading (Boos *et al.*, 2000; Green, 2001). Hens were not separated from the brood for more than 10 minutes. Once a week, ducklings were weighed (± 1g) and immediately released in the pen. Ducklings were caught as groups of 3 or less so that the female was always in contact with some of her brood.

*Hormone assays*

Hormonal plasma levels were determined by radio-immunoassay. Prolactin concentration was determined at CEBC in Chizé (France) using an avian antibody for prolactin (see Lormée *et al.*, 1999, 2000). Testosterone and corticosterone assays were made in our laboratory using $^{125}$I RIA kit double antibody from ICN Biomedicals (http:/www.icnbiomed.com). All plasma samples were assayed at the same time in duplicate to eliminate inter-assay variation. The intra-assay coefficient of variation was 5.5% (n=6 duplicates) and 2.0% (n=8 duplicates) for prolactin and corticosterone respectively (values could not be determined for testosterone because of detection level, see results section).

*Behavioural observations*

Females and ducklings were observed according to the focal sampling method (Altmann, 1974) five times a week by the same observer (A. C.). On the days blood was sampled, no other observations except the recording of vocalization emitted by the hen at release, were performed. Each focal observation lasted 40 min and was carried out between 15:50 and 19:30. From preliminary observations conducted in the field one year before, this period corresponded to maximum daytime activity (Carrière, unpublished data). Observations were randomized weekly so that a subject observed the first day at 15:50 was observed at, for example, 18:50 the following day. Observations started one day after hatching, i.e. when ducklings had left the nest (Kear, 1970). Behavioural monitoring was carried out over 9 weeks, until ducklings fledged (Gollop and Marshall, 1954).

The observer was outside of the pen in an opaque black cabin so that she could not be seen by the ducks. The different behaviours displayed by the animals were monitored through a Plexiglas plate covered with two-way mirror film. The time and duration of each behaviour displayed by the ducklings and the female were recorded with a stopwatch (± 1s).

*Behavioural units*

One year before the present study, a specific ethogram was established in the field on an independent group of free-ranging mallards. Behaviours were classified into four principal categories:

**Self-maintenance behaviours** of the female:

- Foraging: taking food items both in water and on the ground.
- Locomotion: walking and flying.
- Swimming: locomotion in water.
- Preening: cleaning the plumage with the bill or immersion of the head and the neck followed by a rapid raise of the body to sprinkle the back.
- Resting: head placed on the back or the bill under the scapulars.

**Intensity of the hen-brood bond**:

Close proximity i.e. body contacts and distance (<1 m) between the hen and the offspring (duration in s).

Hen being distant (> 1 m) from the ducklings (duration in s).

Keeping proximity i.e. the number of times (per hour) the hen joins the ducklings or the ducklings join the hen.

Disrupting proximity i.e. the number of times (per hour) the hen moves away (> 1 m) from the ducklings or the ducklings move away from the hen.

- Leadership: the number of times (per hour) the hen follows ducklings or

the number of times (per hour) the ducklings follow their mother.

**Vigilance of the hen**:

- Alert behaviours: the hen raises her head or inclines the head with a stretched neck.

- "Motionless Awake-Upright" behaviours: the hen remains upright without moving and observes its surrounds.

**Agonistic behaviours:**

Avoidance: one animal moves away when another one approaches.

Pecking: one animal pecks another one.

These behaviours were recorded whenever they occurred between the hen and offspring or between the ducklings. In addition, to evaluate the strength of cohesion between the ducklings, gathering or dispersion were also noted. The ducklings were considered to be gathered when the distance between them did not exceed 50 cm, and otherwise as dispersed.

In fact, each of these categories of behaviours will enable the determination of the intensity of the parental care each week. If the hen is mainly involved in brood care, she will have less time to care about herself. This probably means that self-maintenance behaviours will vary through the rearing period, being less frequent in the first weeks after hatching and more frequent close to fledging. The intensity of the hen-brood bond will indicate how much the hen takes care of the ducklings by keeping close proximity with them. When parental care decreases, then the hen will probably stay more distant from offspring. Vigilance behaviours are involved in the protection of ducklings all the time since they will be vulnerable to predators. Lastly, agonistic behaviours are the sign of an increase in the independence of the ducklings, which implies that the hen could decrease her attention towards them.

*Statistical analysis*

The duration of behaviours such as foraging, preening, alert, "Motionless Awake-Upright", the distant or contact duration between the female and the ducklings and the gathering/dispersion duration of the ducklings were timed in s and summed on a weekly basis. The duration of observations each week and for each family was equal (5 x 40 min). Specific behaviours such as leadership, maintenance of the proximity when the female or the ducklings move nearer or further away, pecking and avoidance were expressed in frequencies (n/hour) and averaged on a weekly basis. Average body mass and hormone concentrations were calculated on a weekly basis.

Weekly differences were tested with a one-way ANOVA for repeated measurements followed by a post-ANOVA Holm-Sidak test for normally distributed data and after checking for homogeneity of variance (Sigmastat 3.0, SPSS Software). When normality and homogeneity of variance failed, an ANOVA on ranks for repeated measures (Friedman non-parametric test) was performed.

A principal component analysis (PCA) based on the correlation matrix (Minitab 13 software) was carried out on all behavioural parameters displayed by each female over the 9 weeks post-hatching period. Scores resulting from the first axis were used as an index of the intensity of parental care. Parental care was considered as completed when the parental care index remained negative. ANCOVA using weeks as a covariable was used to test for differences of parental care behaviour index and hormone concentrations between females.

Least squares correlations were performed by using Sigma Plot SPSS (7.0).

Values provided are means ± SE. All statistical tests are two-tailed, and probability levels < 0.05 were considered as significant.

## RESULTS

*Timing of breeding and moulting and changes in body mass.*

Hatching occurred between April 15$^{th}$ and June 23$^{th}$ (Fig. 1) with brood sizes ranging from 2 to 14 ducklings. All chicks fledged during week 9 post-hatching.

For the earliest breeding female, moulting of body feathers began on 21$^{th}$ May and between 29$^{th}$ June and 7$^{th}$ July for the others. In all hens, this moult started before offspring fledged but in the three latest breeding females (hatching in June), it was initiated before the end of the parental care phase (Fig. 1). Moulting of wing feathers started between mid-July and early September and well after parental care behaviour had ended.

Body mass of females did not change over post-hatching time ($F_{8,46}$ = 1.40, $p$ = 0.19). Mean body mass of ducklings increased with age according to a sigmoid curve, body mass gain being close to zero at fledging. At week 6 post-hatching, i.e. when parental care was resumed (see below), body mass of ducklings reached 84.2 ± 9.6 % of the hen's body mass. At fledging, ducklings weighed 95.2 ± 0.6% of the hen's body mass.

*Behaviour*

Nine of the 22 behavioural parameters recorded changed significantly ($p$ < 0.05) from hatching to fledging (Table 1). The time spent by the hen resting, preening and being more than 1 m from ducklings as well as the frequency of ducklings pecked by the female increased by 3, 3, 6 and 20 fold respectively during the post-hatching period (Figs. 2a-d). Conversely, the leadership behaviour of the female and of ducklings as well as the time spent by the hen in close proximity with the ducklings decreased with the post-hatching time reaching low values by week 6 (Figs. 3a-c). The frequency of duckling approaches towards the hen and the duration of gathering among the ducklings decreased after week 1 and 6, respectively (Figs. 3d,e).

Alert behaviour of the female tended to change significantly within the 9 weeks post-hatching ($F_{8,46} = 1.90$, $p = 0.08$). It remained at a high level until week 4 post-hatching (range = $466 \pm 69$ to $568 \pm 93$ s/h), then decreased to values ranging from $444 \pm 129$ to $234 \pm 57$ s/h).

From the first week, at the moment of its release, the hen always vocalized. This behaviour ceased after week 6 post-hatching.

The first axis of the PCA, representing the parental care index, accounted for 50% of the variation of all the behavioural variables displayed by the females. It was negatively correlated with sleeping, preening, distance (>1 m) and female pecking (variable loads between -0.42 and -0.25), and positively correlated with contact, female leadership and following (variable loads between 0.34 and 0.40). This parental care index changed over the post-hatching period ($F_{8,46} = 16.34$, $p < 0.001$). It decreased linearly with time post-hatching (Fig. 4a), the mean reaching negative values from week 5 (individual range: week 4 to 6). There was no difference in parental index pattern between hens ($F_{6,47} = 0.09$, NS). Parental care duration and strength were not related to the brood size nor to the hatching date ($F_{1,6} < 1.48$, NS). Mean parental care index decreased together with duckling body mass ($r^2 = 0.98$, $F_{1,8} = 271.04$, $p < 0.001$)

### *Hormones*

Plasma testosterone concentration always remained under the threshold of detection (0.1 ng/ml). Plasma corticosterone concentrations ranged between $33.2 \pm 7.7$ and $53.7 \pm 8.0$ ng/ml and did not change over time post-hatching ($F_{12,62} = 0.88$, NS).

According to post-hatching time, prolactinemia did not differ between females ($F_{6,47} = 0.97$, NS) but changed significantly ($F_{12,70} = 4.92$, $p < 0.001$) over time during the first 13 weeks. Prolactinemia was maximal during the first week ($49.7 \pm 2.4$ ng/ml), and significantly different from values at weeks 7 to 13 ($t_{12} > 3.51$, $p < 0.001$). High levels ($34.2 \pm 5.8$ to $40.3 \pm 3.9$ ng/ml) were maintained until week 6 (Fig. 4b). Plasma prolactin concentrations were then depressed by 36% on average and remained unchanged until week 13, ranging from $21.5 \pm 4.8$ to $30.9 \pm 5.3$ ng/ml.

Prolactin concentrations of the first 6 weeks post-hatching, i.e. corresponding to the parental care phase, were not related to date (julian days, $F_{1,40} = 0.81$, NS), but values from week 7 significantly decreased with date ($r^2 = 0.46$, $F_{1,73} = 63.26$, $p < 0.0001$). In all females, the lowest plasma levels of the reproductive season ($11.4 \pm 0.7$ ng/ml) were reached between the 1st and 18th of August, independently of the hatching date, with inter- and intra-individual variability in the plasma prolactin concentration being reduced five-fold.

*Hormones and behaviour*

Mean weekly values of the parental care index were related to the mean weekly prolactin concentrations according to a sigmoide curve ($r^2 = 0.81$, $F_{3,8} = 7.03$, $p < 0.05$, Fig. 5). The shape of the relationship between the two parameters was similar to that of a dose-response curve, the index of parental care increasing abruptly for plasma concentrations of prolactin ranging between 34 and 40 ng/ml. Logarithmic transformed mean prolactinemia and duckling body mass were negatively correlated ($r^2 = 0.60$, $F_{1,8} = 13.07$, $p < 0.01$). There was no relationship between prolactin and corticosterone concentrations ($F_{1,12} = 1.14$, NS).

# **DISCUSSION**

From the study two major results emerged: 1) there is a progressive decline of parental care during the 6 weeks post-hatching, followed by a non-parental care phase, until fledging, 2) there is a positive relationship between this parental care behaviour and prolactinemia that remains at a high level for several weeks after hatching.

**Behaviour**

Overall, our data obtained in laboratory conditions give new and fundamental evidence to confirm and understand those originating from field observations in ducks and in other precocial species i.e. the degree of female attentiveness toward her brood is the highest within the first two weeks after hatching (Kear, 1970; Afton and Paulus, 1992; Sherry, 1981; Richard-Yris *et al.*, 1988) before brood abandonment after 6-7 weeks post-hatching (Ringelman *et al.*, 1982; Talent *et al.*, 1983). However, for hatching occurring between mid-April and late June, we show that the strength and duration of parental care behaviour 1) was not enhanced by the fact that hens and their brood were maintained in close, forced proximity for up to 13 weeks, 2) was not lower in late than in early breeding females and 3) did not support the *brood-size hypothesis* which suggests that females attending larger broods provide more care to their offsprings (see Pöysä *et al.*, 1997 for a review). This latter result is in accordance with that on brood abandonment fates reported for other dabbling ducks (Gendron and Clark, 2000) and with that reported by Pietz and Buhl (1999) who found no relation between the hen behaviour and brood size in wild free ranging mallards. Overall, we agree with the model developed by Lazarus and Inglis (1986) expecting that, especially in precocial birds, parental investment remains independent from brood size if the investment benefits all ducklings simultaneously and if predators do not take the entire brood when they attack.

The strength of parental care behaviour has been evaluated in some precocial species from the alert behaviour of the female, based on the fact that in species like pink-footed geese, *Anser brachyrhynchus*, and in wild free-ranging mallards, parents rearing a brood spent much more time being alert than those without young (Lazarus and Inglis, 1978; Pietz and Buhl, 1999). From our data this behaviour was observed with the highest frequency during the four weeks post-hatching and tended to decrease afterwards, but non-significantly. In agreement with this latter observation, changes in time spent alert according to the brood age gave controversial results among duck species (see Pietz and Buhl, 1999 for a review). This suggests that the analysis of the alert behaviour alone (or more generally of a single behaviour) is not sufficient to evaluate precisely the strength of parental care. Indeed, of the 22 behavioural parameters we recorded, 40% (pertaining to 3 of the 4 behavioural categories) were significantly modified during the rearing period. In agreement with the leadership character of hen and ducklings, females spent most of their time being close to the brood during the first 5 weeks post-hatching. This was at the expense of the time allocated to self-maintenance behaviours such as preening and resting. From week 6 post-hatching, females clearly expressed an agonistic behaviour toward the brood and the duration of their self-maintenance behaviours increased. At that time the frequency of approach attempts by the brood towards the female was also lowered and, interestingly, the ducklings gathered less. Altogether, these behaviours that developed during the non-parental care period suggest a kind of reciprocal intolerance between the different individuals of the hen-brood unit. To sum up we show here that the hen-brood bond as well as its disruption (here at week 6 post-hatching) appears to be a complex phenomenon requiring a complete behavioural analysis to be fully characterized.

In birds such as ducks, females have to moult before the fall migration. This could lead to possible energy and nutritional conflicts with brood rearing at least in late breeders (Ringelman *et al.*, 1982; Hohman *et al.*, 1992). In our study, body moult (along the sides and flanks) that begun globally in early July, whatever the age of the brood, did not impair the parental care behaviour, and vice versa. Conversely, the intense moult of flight feathers was always initiated after the disruption of the parental care period. This latter result is consistent with the idea of conflicts in nutritional requirement between rearing young and wing moulting (Hohman *et al.*, 1992).

**Prolactin and parental care.**

In precocial birds including ducks, prolactin is involved in parental care and nest attentiveness during incubation (Goldsmith, 1991; Hall, 1991; Criscuolo *et al.*, 2002). In female mallards, mean circulating prolactin concentrations have been shown to increase from about 30 ng/ml to 50-55 ng/ml throughout the course of incubation before dropping sharply immediately after hatching (Goldsmith and Williams, 1980; Bluhm 1992; Hall, 1987). Thus, conversely to altricial species, this hormone was generally considered not to be involved in parental care during young-rearing in precocial birds (see Goldsmith, 1991 for a review). However, results on the Spotted Sandpiper, *Actidis macularia*, the Wilson's Phalarope, *Phalaropus tricolour*, the Bar-headed Goose, *Anser indicus*, the Common Eider and the Bantam hen, *Gallus domesticus*, mitigate this assumption (Dittami, 1981; Oring *et al.*, 1986, 1988; Sharp *et al.*, 1988; Criscuolo *et al.*, 2002). In those precocial species, although prolactin concentration is set at a lower level than during incubation, it remains higher than in adults not rearing young. This fact suggests that this hormone is likely involved in parental care after hatching (Criscuolo *et al.*, 2002). By analysing concomitantly the behaviour and the hormonal profile of the hen we show here that in a precocial species, a well-defined relationship between parental behaviour and prolactin release was present throughout the duckling rearing period. This decline of prolactinemia concomitant to the hen-brood bond disruption, gives strong arguments for a role of prolactin in promoting and/or maintaining post-hatching parental care as demonstrated for altricial species. Moreover, the index of parental care was correlated with prolactin concentrations according to a sigmoide curve (Fig. 5), suggesting that this behaviour may only be stimulated once plasma prolactin concentrations exceed a definite threshold level. Additionally, the contribution of other factors such as the activity and the growth of the ducklings may not be excluded. External stimuli are important factors stimulating prolactin secretion which in turn increase readiness to brood care. In their experiments of nest deprivation Sharp *et al*. (1988) showed a decline in plasma prolactin in the hen when not given chicks, suggesting that chicks may stimulate or at least maintain to some extent prolactin levels for some period of time after hatching. In the same way, changes in duckling growth and behaviour could intervene in the control of prolactin secretion and consequently contribute to stimulate the ending of parental care. Firstly, the mean body mass of the six week old ducklings was about 84% of that of the females and, at this age their morphological aspect (size and plumage) is close to that of the hen (Gollop and Marshall, 1954, Anonymous, 1982). Secondly, considering that ducklings become independent from week 6 by remaining distant from the hen, together with the female intolerance towards ducklings, suggests that the activity changes of ducklings may enhance hen-brood bond disruption through the prolactin secretion pathway (see also

Sharp *et al.*, 1988). Such a process would be adaptive since, at this age, ducklings may become serious food competitors according to the high energy and nutritional needs of the female (see Ringelman *et al.*, 1992).

Although, the release of prolactin is strongly proximately controlled (Silverin and Goldsmith, 1984; Hector and Goldsmith, 1985; Goldsmith, 1991), prolactinemia also partly depends on an endogenously timed mechanism photoperiodically driven (Hector and Goldsmith, 1985; Bluhm, 1992; Garcia *et al.*, 1996; Lormée *et al.*, 1999), such that its levels can, to some extent, be associated with date (Bluhm *et al.*, 1989; Sockman *et al.* 2004). Nevertheless, our data show that during the post-hatching reproductive phase, there was no significant influence of date on the parental care associated prolactinemia pattern. This reinforces the idea that behavioural changes together with duckling growth contributed more than date to the variation of prolactin concentration, at least when hatching occurs from mid-April to late June. However, date had a high influence on prolactinemia after the hen brood-bond disruption. This suggests that the maintenance of a low prolactinemia ($11.4 \pm 0.7$ ng/ml) at the end of the breeding period, whatever hatching date, could correspond to the return at a basal level after a spontaneous ultimately controlled decline of prolactin secretion. Such a decline would be adaptive relatively to the physiological changes the hen is subjected before migration (wing moulting, body fuel and muscles mass adjustments Hohman *et al.*, 1992). As proposed for the Common Eider (Criscuolo *et al.*, 2002), it might therefore be supposed from our data that parental care could no more be promoted or maintained after a basal threshold level has been reached. Nevertheless, we lack sufficient data to demonstrate whether the photoperiod could have a major influence on prolactinemia and the associated post-hatching parental care phase in more late nesting females, i.e. leading to a precocious abandonment of the offspring (Hector and Goldsmith, 1985).

**CONCLUSION**

Under *ad libitum* food supply and captive conditions, we showed that in mallards, a precocial bird, there is a link between prolactin concentration and the parental care behaviour after hatching. This could help to better understand life-history traits about the reproductive investment of the female, which under different environmental conditions, has to match, on the one hand its own nutritional and physiological constraints, and on the other hand protection toward the ducklings until they are able to survive by their own.

**Acknowledgments.**

Financial support to achieve this study was provided by a National Fund for Biological Research on Game and Wildlife Species. We thank Emilie Rauer, Claire Montemont, Alexis Baillis, François Auroy and Philippe Ivanic for technical assistance in blood sampling. Authors are also grateful to André Lacroix (CEBC Chizé) and to Eliane Mioskowski (DEPE) for their technical support in hormone assays, to Dr. A.-S. El Ahmadi for his advises in statistical analyses, to Dr. René Groscolas for his comments on earlier draft, and to Joanna Monroe-Lignot and Lewis Halsey who advised on the English in earlier drafts of the manuscript.

**Figure**

**Fig. 1**. Daylight duration (large dots) for the study area (Strasbourg, 7°45E, 48°35N) and time-course of breeding and moulting for the 7 rearing mallard hens. Periods with (═══) and without (───) parental care are indicated with solid lines whereas the post-fledging period is represented with a dashed line. The times of hatching (H) and fledging (F) as well as the onset of body moult (BM) and wing moult (WM) are also indicated. Brood size is indicated in brackets. The black arrow shows the summer solstice.

**Fig. 2.** Changes in mean (± SE) sleeping (a), preening (b) and distance > 1m (c) durations, and in pecking frequency (d) of hen mallards between hatching and fledging. All changes were significant at $p < 0.05$ (see Table 1).

**Fig. 3.** Changes in mean (± SE) leadership frequencies of hen (a) and duckling (b) mallards, duration of close contact between the hen and the ducklings (c), frequency of duckling approaches towards hen (d) and duration of duckling gathering (e) from hatching to fledging. All changes were significant at $p < 0.05$ (see Table 1).

**Fig. 4.** Changes in the mean (± SE) index of parental care behaviour (a) and plasma prolactin concentration in rearing female mallards over 9 weeks post-hatching. Prolactin values for weeks 10 to 13 (21.5 ± 4.8 to 24.0 ± 6.0) are not shown. Changes were significant at $p < 0.001$ (see text for details).

**Fig. 5.** Relationship between plasma prolactin concentration of mallard hens and parental care behaviour index. Each point represents the mean for each of the 9 weeks post-hatching during which behaviour was studied. One female exhibited a transient low prolactin value at week 4 (i.e. 14.6 ng/ml, 3 times lower than for weeks 1 to 3 and 5, 6). This abnormal value was removed from this analysis.

**Table**

**Table 1.** Variation of behavioural parameters related to the hen-duckling bond during the 9 weeks post-hatching in the mallard ($\chi^2$ indication is for the non-parametric Friedman test). Bold font corresponds to parameters that changed significantly over time post-hatching ($p < 0.05$).

|  | Behavioural parameters | Significance level |
|---|---|---|
| Hen behaviours | **Sleeping** | $\chi^2_8 = 17.34 ; p = 0.027$ |
| | **Preening** | $F_{8,46} = 2.26 ; p = 0.040$ |
| | **Close proximity** | $F_{8,46} = 7.68 ; p < 0.001$ |
| | **Distance > 1 meter** | $F_{8,46} = 5.99 ; p < 0.001$ |
| | **Female leadership (female leads the ducklings)** | $F_{8,46} = 2.43 ; p = 0.028$ |
| | **Duckling leadership (female follows the ducklings)** | $\chi^2_8 = 21.37 ; p < 0.006$ |
| | **Female pecking** | $\chi^2_8 = 16.23 ; p = 0.039$ |
| | "Motionless Awake-Upright" | $F_{8,46} = 1.07 ; p = 0.399$ |
| | Survey | $F_{8,46} = 1.90 ; p = 0.083$ |
| | Feeding | $\chi^2_8 = 5.73 ; p = 0.677$ |
| | Vocalization | $\chi^2_8 = 6.42 ; p = 0.600$ |
| | Locomotion | $F_{8,46} = 0.69 ; p = 0.696$ |
| | IEC | $\chi^2_8 = 10.88 ; p = 0.209$ |
| | Swimming | $\chi^2_8 = 13.33 ; p = 0.101$ |
| | Hen rejoins her ducklings | $F_{8,46} = 1.17 ; p = 0.339$ |
| | Hen avoids her ducklings | $F_{8,46} = 1.18 ; p = 0.329$ |
| | Hen leaves her ducklings | $F_{8,46} = 0.30 ; p = 0.964$ |
| Duckling behaviours | **Ducklings approach the hen** | $F_{8,46} = 2.95 ; p = 0.010$ |
| | **Duckling gathering** | $F_{8,46} = 6.01 ; p < 0.001$ |
| | Ducklings leave the hen | $F_{8,46} = 1.90 ; p = 0.085$ |
| | Pecking of ducklings toward the hen | $F_{8,46} = 1.71 ; p = 0.122$ |
| | Ducklings avoid the hen | $F_{8,46} = 1.54 ; p = 0.171$ |

Figure

Boos et al. Post-hatching parental care behaviour and hormonal status in a precocial bird.

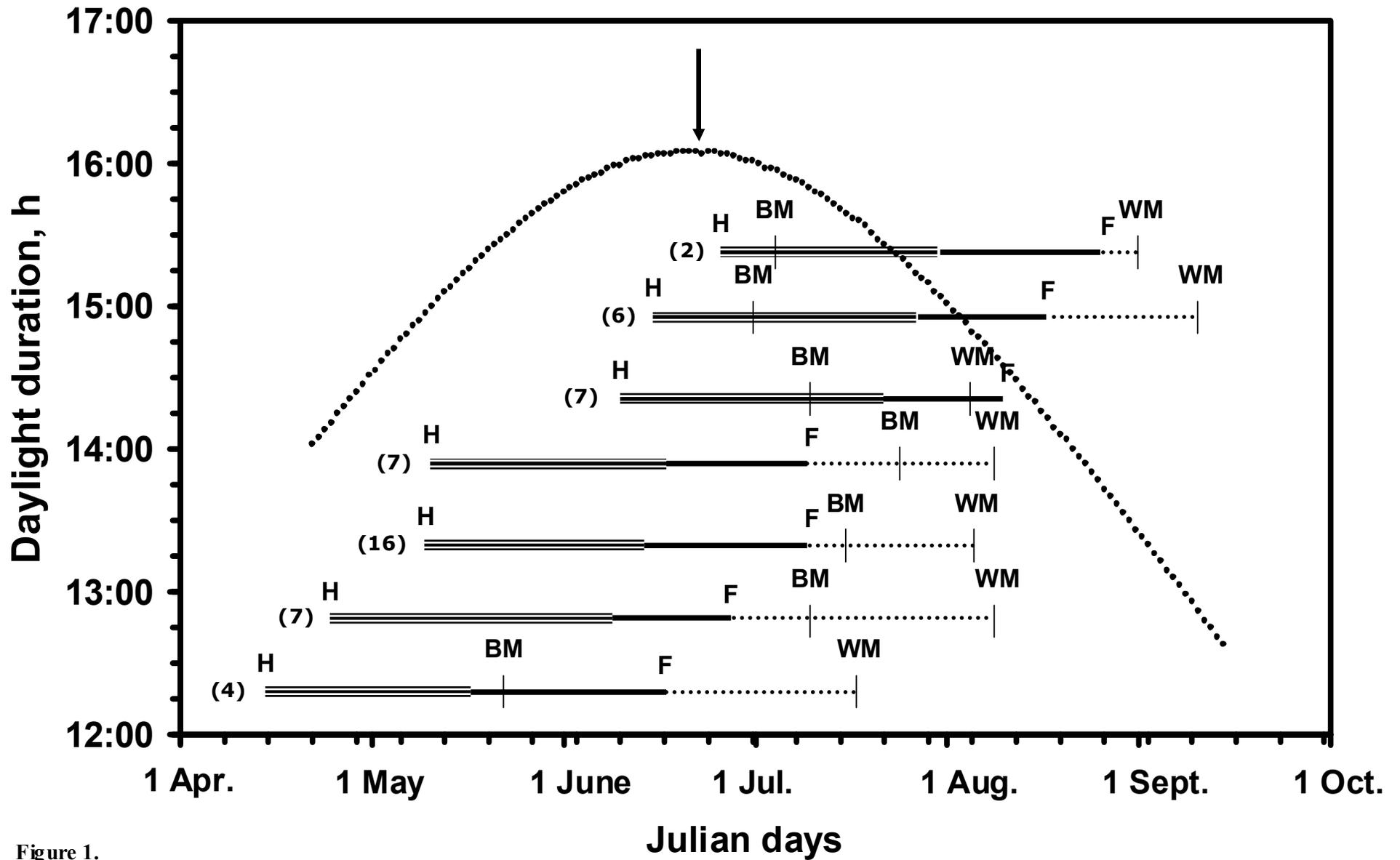

**Figure 1.**

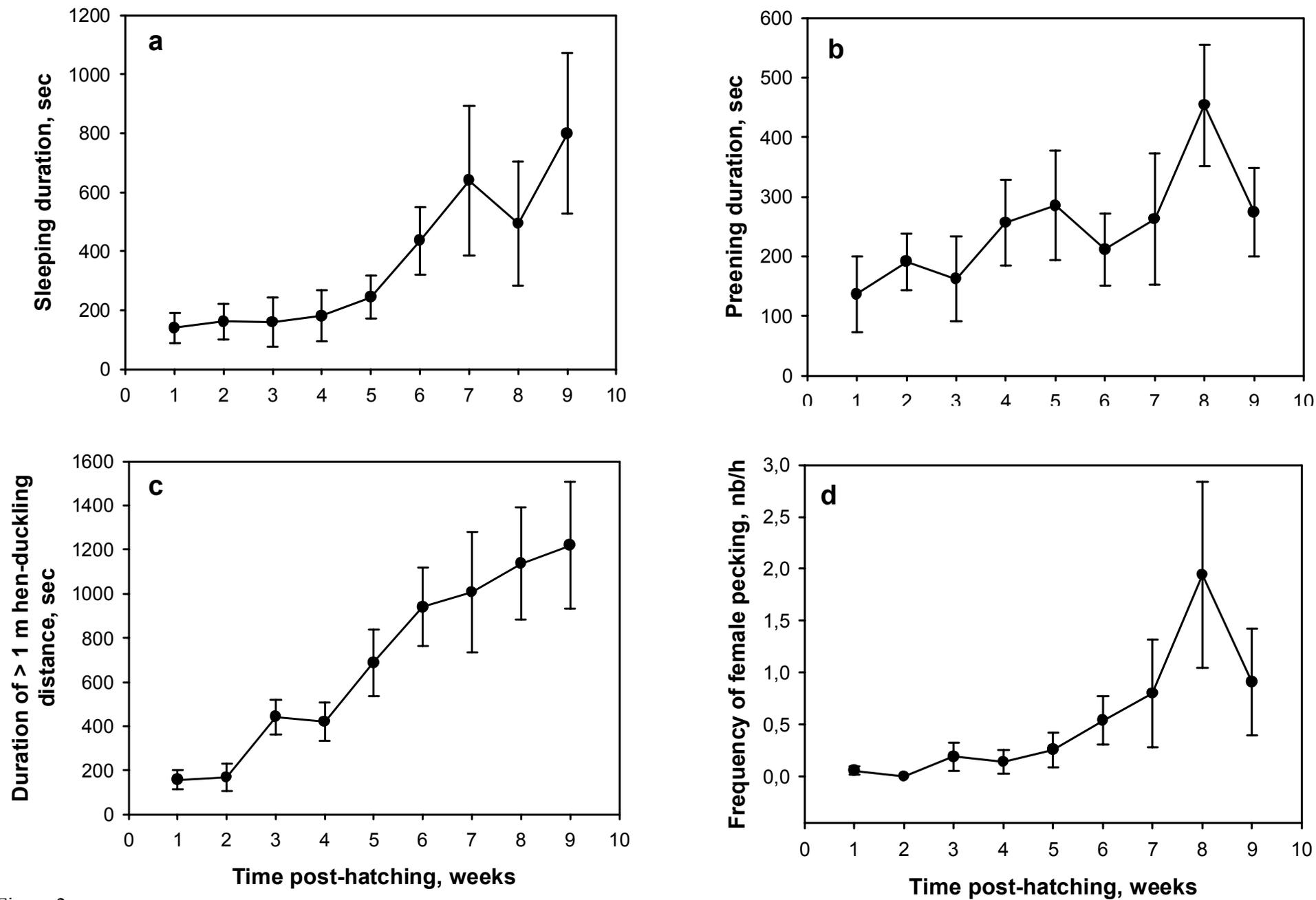

Figure 2.

Figure

Boos et al. Post-hatching parental care behaviour and hormonal status in a precocial bird.

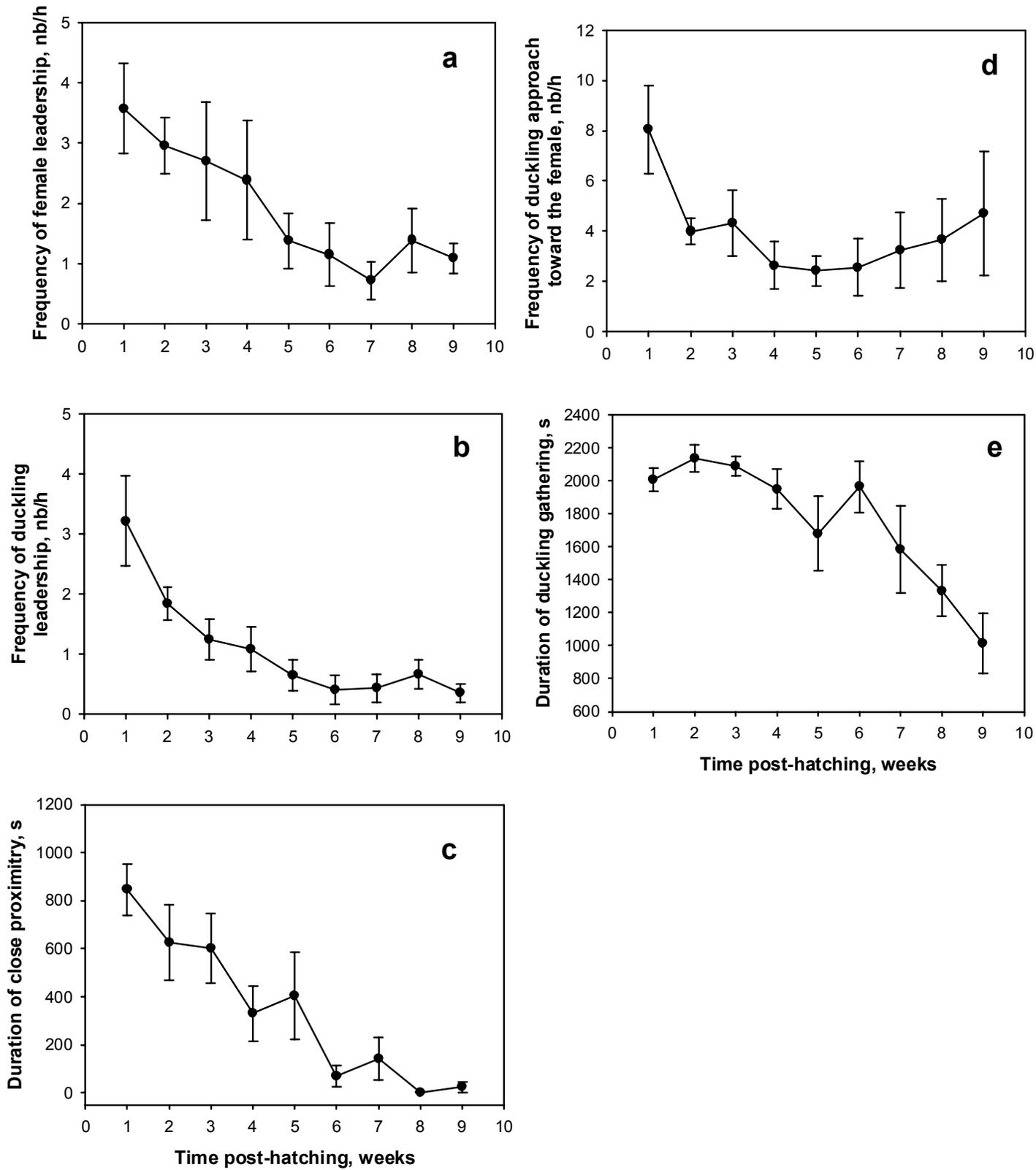

**Figure 3.**

Boos et al. Post-hatching parental care behaviour and hormonal status in a precocial bird.

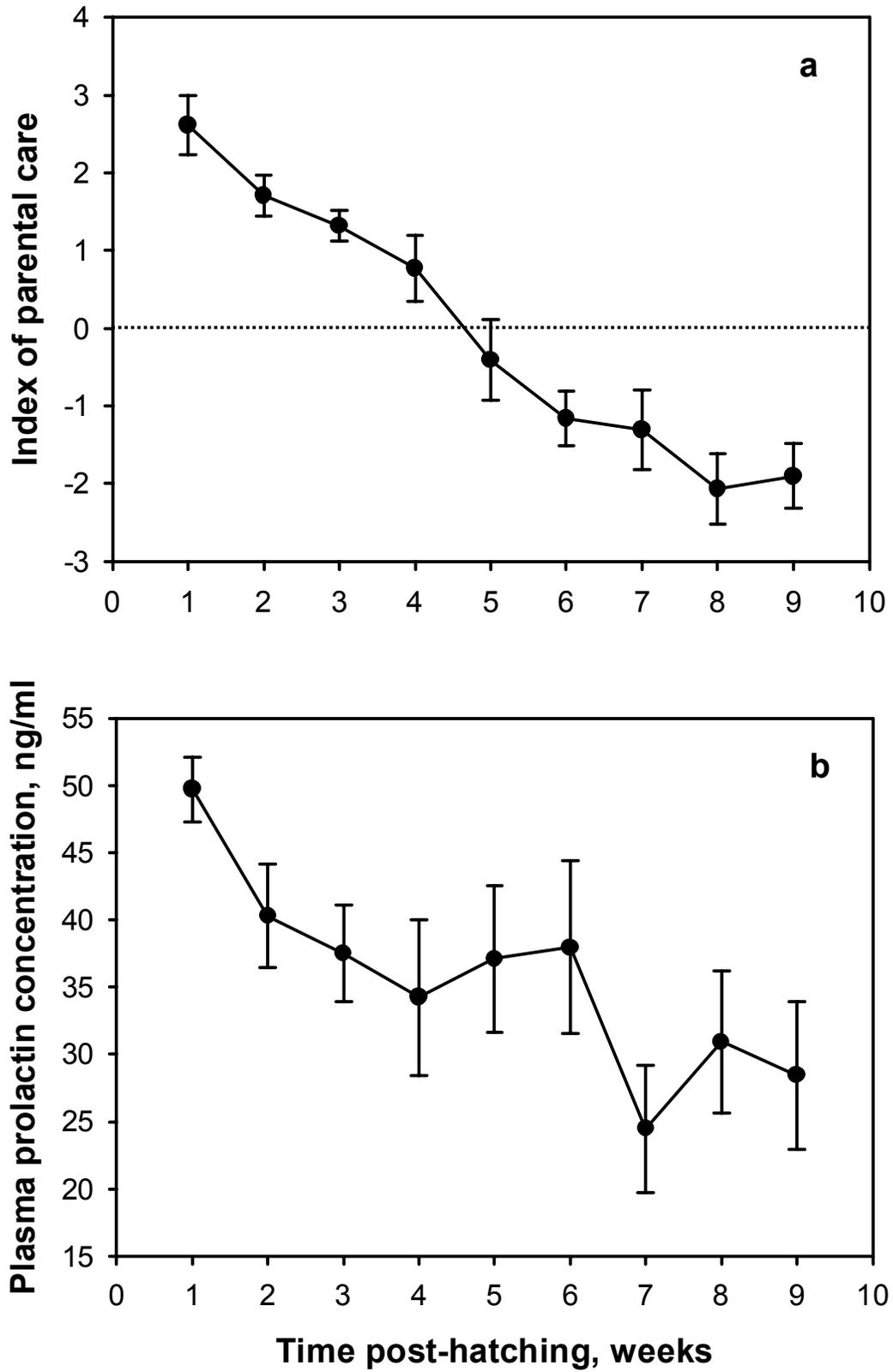

**Figure 4.**

Boos et al. Post-hatching parental care behaviour and hormonal status in a precocial bird.

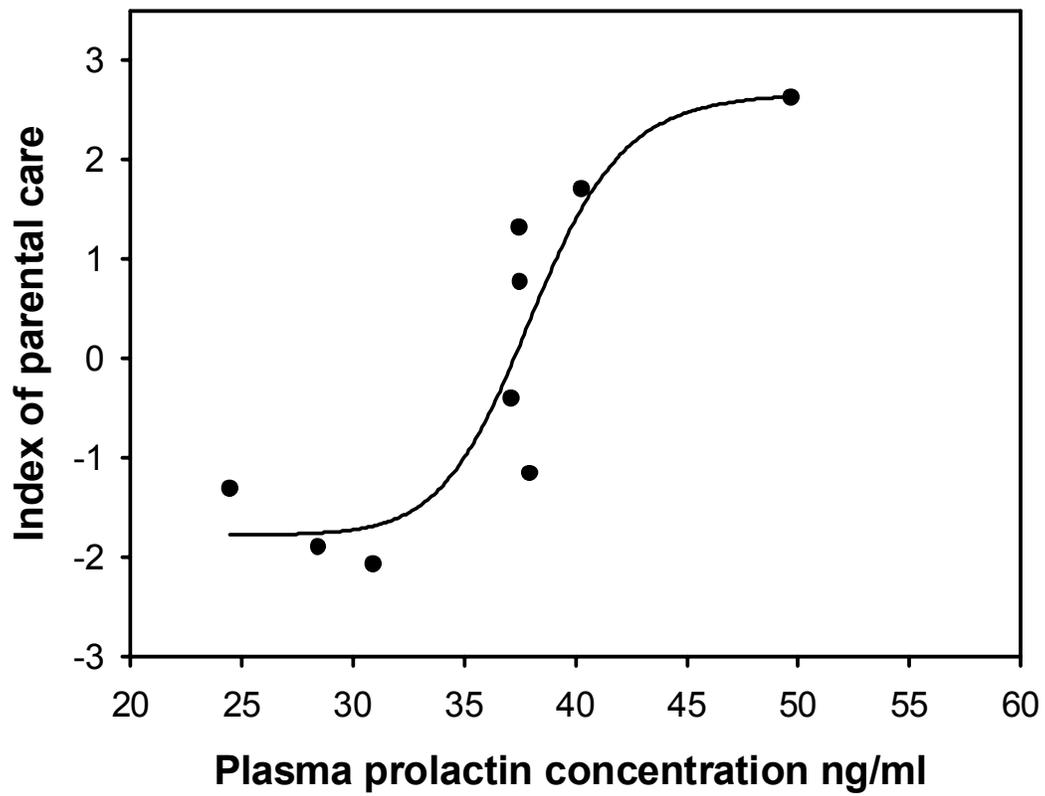

**Figure 5.**